\documentclass[11pt]{article}

\usepackage[left=2cm, right=2cm, top=2.5cm, bottom=2.5cm]{geometry}
\usepackage[x11names]{xcolor}
\usepackage{fancyhdr, amssymb, cancel, amsmath, graphicx, pgfplots, tikz, enumerate}

\newcommand{\stylecolor}{black}

\usepackage[colorlinks=true, urlcolor=violet, linkcolor=blue, citecolor=red, hyperindex=true, linktocpage=true]{hyperref}

\pgfplotsset{samples=200}

\usepackage[explicit]{titlesec}

\newcommand*\sectionlabel{}
\titleformat{\section}
  {\gdef\sectionlabel{}
   \Large\bfseries\scshape}
  {\gdef\sectionlabel{\thesection. }}{0pt}
  {\begin{tikzpicture}[remember picture,overlay]
	\draw (-0.2, 0) node[right] {\textsf{\sectionlabel#1}};
	\draw[thick] (0, -0.4) -- (\textwidth, -0.4);
       \end{tikzpicture}
  }
\titlespacing*{\section}{0pt}{15pt}{20pt}

\newcommand*\subsectionlabel{}
\titleformat{\subsection}
  {\gdef\subsectionlabel{}
   \large\bfseries\scshape}
  {\gdef\subsectionlabel{\thesubsection.\ \  }}{0pt}
  {\begin{tikzpicture}[remember picture,overlay]
    	\draw (-0.15, 0) node[right] {\textsf{\subsectionlabel#1}};
       \end{tikzpicture}
  }
\titlespacing*{\subsection}{0pt}{10pt}{10pt}

\newcommand*\subsubsectionlabel{}
\titleformat{\subsubsection}
  {\gdef\subsubsectionlabel{}
   \bfseries\scshape}
  {\gdef\subsubsectionlabel{\thesubsubsection.\ \  }}{0pt}
  {\begin{tikzpicture}[remember picture,overlay]
    	\draw (-0.15, 0) node[right] {\textsf{\subsubsectionlabel#1}};
       \end{tikzpicture}
  }
\titlespacing*{\subsubsection}{0pt}{7pt}{7pt}

\pgfplotsset{every axis legend/.append style={at={(1.02,1)},anchor=north west}}

\newcommand{\titletext}{Connecting microscopic physics with the macroscopic properties of materials in introductory physics courses}

\begin{document}

\pagestyle{fancy}
\renewcommand{\headrulewidth}{0pt}
\fancyhead{}

\fancyfoot{}
\fancyfoot[C] {\textsf{\textbf{\thepage}}}

\begin{equation*}
\begin{tikzpicture}
\draw (0.5\textwidth, -3) node[text width = \textwidth] {{\huge \begin{center} \color{\stylecolor} \textsf{\textbf{\titletext}} \end{center}}}; 
\end{tikzpicture}
\end{equation*}
\begin{equation*}
\begin{tikzpicture}
\draw (0.5\textwidth, 0.1) node[text width=\textwidth] {\large \color{black} $\text{\textsf{Andrew Lucas}}$};
\draw (0.5\textwidth, -0.5) node[text width=\textwidth] {\small  \textsf{Department of Physics, Harvard University, Cambridge, MA, USA, 02138}};
\end{tikzpicture}
\end{equation*}
\begin{equation*}
\begin{tikzpicture}
\draw (0.5\textwidth, -6) node[below, text width=0.8\textwidth] {\small  An elementary understanding of the relevant length, mass and energy scales at the molecular level can be used to explain the order of magnitude of material properties such as mass density, latent heat, surface tension, elastic moduli and beyond in an introductory physics course.   These order of magnitude estimates are remarkably easy to derive, and in many instances are the same for many different liquids and solids. This helps students to understand the origin of the zoo of material properties, and to connect molecular physics to the physics of familiar materials like water, metals and plastics.  We also note some simple mechanisms by which material properties can easily vary over many orders of magnitude.};  
\end{tikzpicture}
\end{equation*}
\begin{equation*}
\begin{tikzpicture}
\draw (0, -13.1) node[right, text width=0.5\textwidth] {\texttt{lucas@fas.harvard.edu}};
\draw (\textwidth, -13.1) node[left] {\textsf{\today}};
\end{tikzpicture}
\end{equation*}

\tableofcontents

\section{Introduction}
Engineers and experimental scientists must confront a zoo of material properties.   Even for  a substance like water, the number of important material properties is mind-boggling:  the latent heats of melting and boiling, the specific heat, the electrical and thermal conductivities, the mass density, the compressibility, the speed of sound, etc.   Giant handbooks \cite{Haynes2013} are used to collect these properties, and  typically, teachers simply look up the relevant numbers for various materials, for students to use in lab projects, homework or exams.

Rather remarkably, a large fraction of these parameters can usually be \emph{estimated to within an order of magnitude} using elementary physical ideas which are accessible to freshmen in introductory physics courses, or even advanced high school students.   Whenever these parameters cannot be reliably estimated as such, it is actually a signal that very interesting physics is at play.

The goal of this paper is to point out to teachers of introductory physics the remarkable ease with which most material properties can be estimated from elementary molecular physics.   Not only does this explain of the order(s) of magnitude of so many material properties of solids and liquids  (why do materials made of so drastically different components, arranged in such drastically different ways, all have the same mass density to an order of magnitude or so), but it also helps students to connect the microscopic world of atomic and molecular physics to facts they are familiar with about the everyday world.     As many aspects of the universal behavior of gases are well-explained in statistical mechanics \cite{Reif1965}, we will not discuss this.    The universal orders of magnitude of many of the properties of solids and liquids, however, is not well-explained in any physics book that we have seen, with rare exceptions for a few properties, in some books.   

Not every property can be estimated in this way, and at the end of this paper we will briefly comment on some cases where order of magnitude estimates cannot be made, and explain some reasons why.  One example is the electrical behavior of materials, which depends strongly on the nature and mobility of charge carriers and can vary strongly between materials (e.g., metals vs. insulators) \cite{Ashcroft1976}, though we will not discuss electrical properties of materials further.

\subsection{A Note on the Units of Temperature}
Before beginning, we should make an important point.   We will be working in physicist's units, where temperature $T$ has  the dimensions of energy: \begin{equation}
[T] = [\text{energy}].
\end{equation}(The square brackets will denote dimensions throughout the paper.)   This contrasts with the usual units where $k_{\mathrm{B}}T$ has units of energy, and temperature has a special dimension.   Unlike other units such as length and mass, where there are \emph{distinct physical quantities} associated with length, time, energy, etc., there is no physical content in Boltzmann's constant $k_{\mathrm{B}}$;   it is simply a unit conversion factor.     Teachers should emphasize this fact to students, and that what temperature truly is is a measure of how likely fluctuations in energy are:  (microscopic) energy fluctuations of order $T$ are likely, and fluctuations of order $\gg T$ are very rare.      Room temperature in units of energy is about $T_{\mathrm{room}} = 4\times 10^{-21}$ J.   This is important to understand:  sometimes the energy scale relevant to a problem is temperature.

\section{Microscopic Physics} \label{sec2}
Let us begin with a brief overview of microscopic physics.   Depending on the level of the students, this information can either be derived, or just stated as a fact.   It should not be unreasonable to simply state these facts, since we will use remarkably few facts to obtain a huge amount of information on macroscopic material properties.   Our opinion is that these facts can be stated early on in a course, and some of them (such as the  energy scales of chemical bonds) can be ``derived" later (mostly by similar back-of-the-envelope calculations to this paper) when basic ideas about quantum physics are introduced.

\subsection{Simple Solids and Liquids}
We are all taught from a young age that solids and liquids are made up of atoms.   So let's begin by talking about scales associated to atoms.    For simplicity, let us consider the hydrogen atom, which has mass $m_{\mathrm{H}} \approx 2\times 10^{-27}$ kg.  The energy scale associated with this atom is $E_{\mathrm{H}} \sim 10^{-18}$ J, and the length scale is the Bohr radius, $a_{\mathrm{H}} \approx 5\times 10^{-11}$ m \cite{Atkins2009}.\footnote{We see that SI units are \emph{bad} units to talk about molecular physics.  However, because we want to discuss macroscopic material properties, which for better or worse are often measured in SI units for historical reasons, we will use these units throughout.}

However, we will rarely be thinking about pure hydrogen.    When we talk about solids like ice or salt, made up of atoms bonded together, the appropriate mass scale will change simply by a factor proportional to the number of protons and neutrons (which approximately have the same mass) per atom or molecule.   The appropriate length scale is now the length of a chemical bond, which we can crudely estimate to be twice the Bohr radius:  $a_{\mathrm{B}} \approx 0.1$ nm.   This is amusingly close to 0.12 nm, which is the carbon-carbon (single) bond length (this is one of the most common bonds in nature).    The appropriate energy scale, up to O(1) factors, is unchanged:    the chemical energy required to break a bond is usually about $0.5E_{\mathrm{H}}$ \cite{Atkins2009}.       This is because the same quantum physics is responsible for determining the energy of a chemical bond.     We will often be thinking implicitly of chemical bonds as classical springs, an analogy which is often used in introductory physics texts \cite{Feynman1965b}.

Finally, sometimes we will see that the intermolecular scale, $a_{\mathrm{I}} \approx 0.3$ nm, is relevant.    We expect this to happen when the relative distances between molecules are being increased or decreased as we distort a material.   This is the case when we deal with a liquid such as water (although one should be careful, as water is actually more dense than ice, at standard freezing temperatures).   Students can understand this scale heuristically as well, as we expect  $a_{\mathrm{I}} \gtrsim 2a_{\mathrm{B}}$, as the molecules should be distinct, so thus the distance between molecules should be a bit larger than the length of a chemical bond.   If one does not want to distinguish between $a_{\mathrm{I}}$ and $a_{\mathrm{B}}$, this is not too disastrous, although when volumes are relevant, these factors lead to a change of an order of magnitude, and are important to keep.    Since liquids are interacting only through intermolecular forces,  it turns out that the energy scale is about $E_{\mathrm{I}}\sim E_{\mathrm{H}}/20$; for simplicity, we will usually round this to about $10^{-19}$ J \cite{Atkins2009}.

\begin{figure}[h]
\centering
\includegraphics{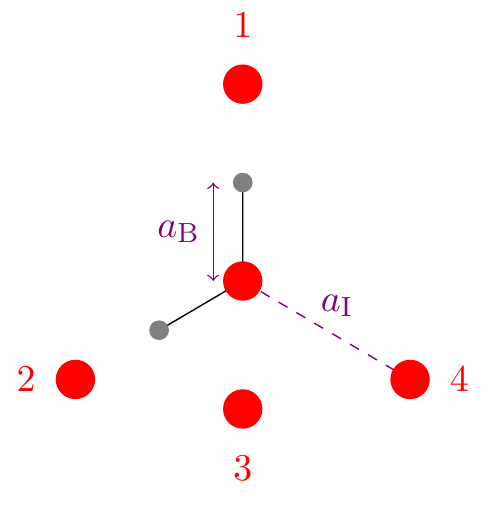}
\caption{Linus Pauling's model of water.   The two hydrogens (small circle) are each oriented towards an external oxygen (big circle).   The intermolecular and chemical bond scales are also shown.}
\label{fig1}
\end{figure}

As a toy example of these ideas, it is helpful to consider liquid water.   Linus Pauling long ago estimated that one can consider water to be approximately in a tetrahedral lattice \cite{Pauling1935, Phillips2012}, where the two hydrogens of $\mathrm{H}_2\mathrm{O}$ are oriented towards any two of the vertices of the tetrahedron, as shown in Figure \ref{fig1}.   The two different scales $a_{\mathrm{B}}$ and $a_{\mathrm{I}}$ are clearly shown as well.   We will use this toy model later in the paper.
\subsection{Polymers}
Let us now discuss a rather different type of material -- polymers.    A polymer is a very long chain like molecule made up of hundreds or thousands of simple units, called monomers;   the properties of these molecules can be  radically different due to their large size.  As we will see, in some cases, polymers have very similar properties to other materials, and in other cases very different properties.   Our goal will be to understand the difference.     Since polymers are built up of bonded atoms, we expect the relevant length scales to be $a_{\mathrm{B}}$ (and $a_{\mathrm{I}}$), and the relevant mass scale to be $m_{\mathrm{H}}$.

Whereas the energy scales in ``simple" materials are quantum energy scales, the energy scale in a polymer is very different.   The key change is that the energetic properties of polymers are often \emph{thermal} in nature, and so the relevant energy scale in a polymer is the temperature $T$.   We can understand this crudely as follows.   The free energy of a polymer is given by $E-TS$, where $E$ is the mechanical energy and $S$ is the entropy.   If each bond is slightly tilted, then we can slowly wind the polymer around, and the choice of the direction of winding gives us entropy, at very little energetic cost.   So at the minimum of free energy, the chemical bonds are not severely stretched, and the free energy is dominated by entropic effects.   In particular, we see that if the free energy is approximately $-TS$, then the only energy scale in the problem is temperature $T$, as entropy is a dimensionless quantity.

\section{Mechanical Properties}\label{sec4}

\subsection{Mass Density}
We begin with a rather elementary consideration, which is the mass density of a solid or fluid.   Students may be rather surprised that while gases are very light:  $\rho_{\mathrm{air}} \sim 1$ kg/$\mathrm{m}^3$;  solids and fluids are comparably ``heavy":   $\rho \sim 10^3$ kg/$\mathrm{m}^3$.    The reason for this is very simple:   unlike a gas, the constituents of a fluid or solid strongly interact with each other, and try to get as close to one another as possible.  

Indeed, we expect fundamental limits on the density of matter, simply because we can only squeeze so many atoms into a given space.\footnote{This is until we reach the limit where atoms themselves are ill-defined, such as in a neutron star or a quark-gluon plasma.}    By dimensional analysis, this follows naturally:  a mass density corresponds to a mass per volume: \begin{equation}
[\rho] = \frac{[\text{mass}]}{\text{[volume]}}.
\end{equation}For simplicity, let us suppose we had matter made up only of hydrogen atoms.   Using the mass $m_{\mathrm{H}}$ and the length scale $a_{\mathrm{H}}$ associated to hydrogen, we find \begin{equation}
\rho_{\mathrm{max}} \sim \frac{m_{\mathrm{H}}}{a_{\mathrm{H}}^3} \sim 1.6\times 10^4 \; \frac{\mathrm{kg}}{\mathrm{m}^3}. 
\end{equation}
Crudely, we can posit that ``no matter on Earth should exist which is more dense than $\rho_{\mathrm{max}}$" if it is made up of well-defined atoms.   Remarkably, the most massive known material in the world, osmium, has $\rho\approx2.2\times 10^4$ $\mathrm{kg}/\mathrm{m}^3$, which is embarrassingly close to this crude estimate \cite{Arblaster1989}:  We have not even taken into account any factors associated with the ease of sphere packing, for example.

 If students object to this because we could have used a heavier atom, one can note that (roughly speaking) the size of the electron cloud around an atom grows as $a\sim m^{1/3}$ if one does a more careful calculation (see the Thomas-Fermi equation in \cite{Landau1977}), so our estimate would not change.    We further note that, within an order of magnitude, many crystalline solids have this mass density.

The mass density of liquids is often quite a bit smaller.    This is because the molecules in a liquid are separated a bit further and interact through intermolecular forces.   The water molecule has mass $18m_{\mathrm{H}}\approx 3\times 10^{-26}$ kg; we have seen the appropriate length scale here is $a_{\mathrm{I}}$.   One can thus estimate the density of water as about $1000$ kg/$\mathrm{m}^3$, which is essentially the correct answer.       Intermolecular length scales are important in solids made up of crystals of interacting organic molecules (such as sugar); it is not surprising, therefore, that such solids have similar mass densities to water.

\subsection{Young's Modulus}
Young's modulus for an elastic solid can be defined analogously to Hooke's law, as \begin{equation}
\sigma = E \frac{\delta L}{L}
\end{equation}where $\sigma$ is the stress (force per unit area) applied to an elastic solid of original length $L$, if it has been compressed to a length $L-\delta L$.    By dimensional analysis, one sees that \begin{equation}
[E] = [\text{pressure}] = \frac{[\text{force}]}{[\text{area}]} = \frac{[\text{energy}]}{[\text{volume}]}.
\end{equation}

Let us begin by considering a ``strong" material like diamond.   The relevant length scale here will be the chemical bond length scale $a_{\mathrm{B}}$, since the molecule as a whole is getting squeezed, and the relevant energy scale is that of chemical bonds.  We conclude that \begin{equation}
E \sim \frac{10^{-18} \; \mathrm{J}}{( 10^{-10} \; \mathrm{m})^3} \sim 10^{12} \; \mathrm{Pa}.   \label{eqe}
\end{equation} Indeed, diamond has $E_{\mathrm{diamond}} \approx 1200$ GPa.    Computer simulations have suggested that another carbon-based solid, \emph{carbyne}, may have a Young's modulus of 32000 GPa, an order of magnitude higher \cite{Liu2013}.   While incredibly impressive, it still suggests that the scale of 1000 GPa is, crudely, a fundamental limit on the stiffness of a material, a fact observed in practice \cite{Courtney2000}.     For other crystals like copper (100 GPa) and sapphire (400 GPa), the numbers are a bit lower, but still not smaller by more than an order of magnitude.

Next, we consider a material like glass.   Glass is made up of $\mathrm{SiO}_2$, arranged in an amorphous way.   We might expect that, like in water, the relevant length scale is intermolecular.   A material where the relevant length and energy scales are similar would be wood, made up of complicated proteins packed together.   Although proteins are polymers, here the polymeric nature is not important:  the individual chains are mostly frozen in a ``crystalline" arrangement, and thermal physics is not an important effect.   In both cases, we might expect that we should replace the energy scale in Eq. (\ref{eqe}) with $E_{\mathrm{I}}$, because the crystalline structure is made up of well-defined molecules.    We thus estimate that the Young's modulus will drop to about 50 GPa.   This is heuristically consistent with typical values for  glass ($E_{\mathrm{glass}} \sim 60$ GPa).   For wood, we find $E_{\mathrm{wood}}\sim 10$ GPa, in many instances.   The value for wood is likely a bit lower because intermolecular energy scales begin to play a role as well.   In fact, if we assume that the length scales are $a_{\mathrm{B}}$, but the energy scale is $E_{\mathrm{H}}$, we obtain that $E\sim 30$ GPa, which is a very similar estimate.   It is thus likely that some combination of increasing characteristic energy and length scales is responsible for the smaller Young's modulus of wood.

Let us now see what changes if we work with a polymeric material, such as plastic.    In this case, we expect that classical entropic considerations dominate over the quantum energy of a chemical bond, as we discussed previously.   So our first estimate of the Young's modulus of a polymeric solid would be $E\sim Ta_{\mathrm{B}}^{-3} \sim 4$ GPa.  Polymeric materials have widely varying values of $E$, typically ranging from 10 MPa to 10 GPa \cite{Courtney2000}, and we have made an estimate towards the upper end of this range.  It is not difficult to give qualitative explanations for why some polymers have larger $E$:  for example, in vulcanized (tire) rubber, sulfide bonds between polymer chains dramatically stiffen the material.  In this case, we might expect the dominant contribution to $E$ to once again become due to the squeezing of chemical bonds, and not to entropic effects.   However, there are many polymers which are far less stiff than we predicted.  A simple reason for this is as follows -- we implicitly assumed that the appropriate length scale was $a_{\mathrm{B}}$.  However, the appropriate length scale may be $a_{\mathrm{I}}$ (or larger) for two reasons:  firstly, many monomers are the sizes of small molecules, and secondly, we expect interchain spacings to be typically larger than $a_{\mathrm{B}}$.   Replacing $a_{\mathrm{B}}$ with $a_{\mathrm{I}}$, we find that $E\sim Ta_{\mathrm{I}}^{-3}\sim 100$ MPa, which is much closer to the lower bound on typical polymer Young's moduli.

\subsection{Compressibility of Liquids}
The compressibility $\kappa$ of a liquid is defined as follows:  \begin{equation}
\kappa = - \frac{1}{V} \frac{\partial V}{\partial P}.
\end{equation}By dimensional analysis, we see that $1/\kappa$ has the dimensions of a pressure.    As we argued when discussing mass density, the arguments for scaling behaviors of solids and liquids are in many cases quite similar.   However, for a liquid, where the molecules are not arranged in a rigid crystal lattice, we expect the intermolecular energy and length scales to be relevant; otherwise, the calculation should be identical to the Young's modulus calculation.   So we expect \begin{equation}
\frac{1}{\kappa} \sim \frac{E_{\mathrm{I}}}{a^3_{\mathrm{I}}} \sim 4 \; \mathrm{GPa}  \label{kappa}
\end{equation}   And indeed, this is remarkably the case -- the proper units for compressibility are $\mathrm{GPa}^{-1}$:  for example,  $\kappa_{\mathrm{water}} \approx 0.4 \; \mathrm{GPa}^{-1}$ \cite{Fine1973}.

\subsection{The Speed of Sound}
Another interesting material property of a solid the speed at which sound waves travel.   It is a classic textbook formula that the speed of sound in a material is given by\footnote{This is not quite true -- there are two modes of propagation in a solid \cite{Kinsler2000}.   But qualitatively, it is true;  the two modes have the same speeds up to an O(1) factor.} \begin{equation}
c \sim \sqrt{\frac{E}{\rho}}.
\end{equation}It is not surprising, based on dimensional analysis, that this formula should be true;  we have also done all of the work already to find $c$.   In a crystalline solid, we expect $c\sim 10^{4}$ m/s, which is the correct order of magnitude for almost all materials:  e.g. steel has a (P-wave) sound speed of about 6000 m/s \cite{Kinsler2000}.

In a liquid, we have instead \begin{equation}
c=\frac{1}{\sqrt{\rho\kappa}}.
\end{equation}We find for this estimate that $c\sim 10^3\; \mathrm{m}/\mathrm{s}$, which again is a reasonable estimate for many materials: e.g., 1500 m/s for water \cite{Kinsler2000}.

\section{Thermodynamic Properties}\label{sec6}

\subsection{Latent Heats}
Let us begin by asking how much energy it takes to melt a solid.   This is frequently defined in terms of the latent heat $L$, defined as the ratio of the heat required to melt a solid, to the mass of that solid: \begin{equation}
L  \equiv \frac{Q}{M}.  \label{eq12}
\end{equation}Let us begin by noting that $M=\rho V$, where $V$ is the volume of the solid, and $\rho$ is the mass density, whose microscopic origins we have understood.   To compute $Q$, let us estimate that the entropy of the liquid phase is substantially greater than that in the solid phase -- so much so, in fact, that we can estimate that the solid phase has zero entropy.\footnote{In reality, the entropy changes by an O(1) factor, but the argument below contains all of the essential physics.}   This is a crude assumption, but it is reasonable as a first approximation.   So the heat transfer is then given by\begin{equation}
Q = TS_{\mathrm{liquid}}.  \label{eq13}
\end{equation}The entropy of the liquid phase can be estimated by counting the total number of configurations of the liquid.   It will be helpful to consider the specific case of water, as discussed earlier in Figure \ref{fig1}.   Simple combinatorics gives us that there are 6 possible orientations of the water molecule: \begin{equation}
W_{\mathrm{mol}}  = \left(\begin{array}{ll} 4 \\ 2 \end{array}\right) = \frac{4!}{2!2!} = 6.
\end{equation}So we can estimate that the entropy of the liquid phase is simply given by all possible orientations of these water molecules: \begin{equation}
S_{\mathrm{liquid}} \sim \log \left( {W_{\mathrm{mol}}}^{V/a_{\mathrm{I}}^3} \right) \sim \frac{V}{a_{\mathrm{I}}^3} \log 6.  \label{eq15}
\end{equation}Combining Eqs. (\ref{eq12}), (\ref{eq13}) and (\ref{eq15}) we find that (letting $m$ be the mass of the water molecule) \begin{equation}
L = \frac{T\log 6}{\rho a_{\mathrm{I}}^3} = \frac{T}{m} \log 6 \approx 10^5 \; \frac{\mathrm{J}}{\mathrm{kg}},
\end{equation}which is the same order of magnitude as the empirical result: $L_{\mathrm{water}} \approx 330\; \mathrm{kJ}/\mathrm{kg}$.   Indeed, one finds that the units of $10^5\; \mathrm{J}/\mathrm{kg}$ are the correct units in which to measure latent heats of fusion for a variety of liquid-solid transitions \cite{Atkins2009}.

Similarly to above, let us ask how much energy it would take to vaporize a liquid.   One might crudely try and estimate that if one has to overcome intermolecular forces to vaporize a liquid, we should expect that, for water, e.g.:\begin{equation}
L_{\mathrm{vap}} \sim \frac{Q}{M} \sim \frac{E_{\mathrm{I}}}{m} \sim \frac{10^{-19} \; \mathrm{J}}{10^{-25} \; \mathrm{kg}} \sim 10^6 \; \frac{\mathrm{J}}{\mathrm{kg}}.
\end{equation} The correct values for $L_{\mathrm{vap}}$ are typically about $10^6\; \mathrm{J}/\mathrm{kg}$ \cite{Atkins2009}, remarkably close to this crude estimate.  

A complementary and more thorough discussion of latent heats, appropriate for a slightly more advanced audience, can be found in \cite{bernstein}.   In particular, alternative justifications for why the energy required to melt a solid should be of order $T$ are presented, which  result in the same estimate as we made above.

\subsection{Thermal Expansion Coefficient}
Let us now argue what the thermal expansion coefficients should be.    Thermal expansion coefficients are defined as follows:  let us consider a cube of solid or liquid material with side length $L$.   Then if we heat it up by a small amount, the cube will grow in size according to the formula:\begin{equation}
\alpha = \frac{1}{L} \frac{\partial L}{\partial T}.  \label{eq25}
\end{equation}

To estimate $\alpha$, consider the following simple model.   Suppose that $a$ is the length of the material per chemical bond, and $a_{\mathrm{B}}$ is the chemical bond distance at very low temperatures.   Then, as a function of the dimensionless parameter $u\equiv a/a_{\mathrm{B}}$, we can approximate that the energy of the chemical bond is given by \begin{equation}
E_{\mathrm{bond}}(u) = -E_{\mathrm{H}} V(u),
\end{equation} where $V(u)$ is an O(1) function of an O(1) parameter.   At finite temperature $T$, the expectation value of $u$ is given by \begin{equation}
\langle u\rangle \sim  \int \mathrm{d}u \; u \exp\left[-\frac{E_{\mathrm{bond}}(u)}{T}\right] =  \int \mathrm{d}u \; u \exp\left[-\frac{E_{\mathrm{H}}}{T}V(u)\right] 
\end{equation}An equivalent definition for $\alpha$ is given by $\alpha = \partial \langle u\rangle /\partial T$, because there are so many chemical bonds in a material that the length is proportional to $\langle u\rangle$.   We see that $\langle u\rangle$ is dependent only on the dimensionless ratio $E_{\mathrm{H}}/T$, we conclude that $\alpha \sim 1/E_{\mathrm{H}}$, so long as the $E_{\mathrm{H}}/T$ dependence  is not too harsh.   

If we have a liquid like water, we need to replace $a_{\mathrm{B}}$ with $a_{\mathrm{I}}$, and $E_{\mathrm{H}}$ with $E_{\mathrm{I}}$, in the above derivation.   We see that in this case $\alpha \sim 1/E_{\mathrm{I}}$.    If we have a complex polymeric material where the only energy scale is temperature $T$, we conclude by dimensional analysis that $\alpha \sim 1/T$.

Let us compare this to real materials.     We expect materials where chemical bonds set the energy scale to have $\alpha \sim 10^{18} \; \mathrm{J}^{-1}$.  This is heuristically true, with some large deviations:  e.g., diamond has $\alpha \approx 10^{17} \; \mathrm{J}^{-1}$, whereas copper has $\alpha \approx 10^{18}\;\mathrm{J}^{-1}$ \cite{Atkins2009}.\footnote{This is usually given in experimental units of $\mathrm{K}^{-1}$.  Using Boltzmann's constant, one simply multiplies our numbers by $k_{\mathrm{B}}\sim 10^{-23}\;\mathrm{J}/\mathrm{K}$ to convert.   For example, one finds that diamond has $\alpha \sim 10^{-6} \; \mathrm{K}^{-1}$.}     If intermolecular interactions set the energy scale, we expect $\alpha \sim 10^{19}$ J: water has $\alpha \approx 7\times 10^{18} \; \mathrm{J}^{-1}$, which is reasonable.       Another good check on our insight is whether a polymer, where the effective energy scale should be $T$, has $\alpha \sim T^{-1} \sim 10^{20}\; \mathrm{J}^{-1}$ .      One in fact finds, for a soft plastic, that $\alpha \sim 2\times 10^{19} \; \mathrm{J}^{-1}$, which is a bit lower than our estimate, but still larger than for most other classes of materials.

A more careful test is to note that $\alpha E \sim 1$ (here $E$ is Young's modulus) -- this fact is known to be qualitatively true \cite{Courtney2000}.

\subsection{Surface Tension}
The calculation of the surface tension in between water and a hydrophobic liquid is very similar to the previous calculation of $L$.   This particular calculation is produced in the beautiful textbook \cite{Phillips2012}, and we repeat it here for completeness.

The surface tension at the interface between, say, oil and water, is due to the fact that water cannot hydrogen bond with an oil molecule (hydrocarbon).   Suppose there is an interface of area $A$ between the oil and water.    For each missed opportunity to hydrogen bond, there is a free energy penalty.      In particular, referring to Figure \ref{fig1}, let us suppose that the presence of an oil molecule blocks hydrogen bonding with site 1;  now there are only 3 configurations for the water molecule. The free energy penalty is therefore \begin{equation}
\Delta F = -T\log 3 + T\log 6 = T\log 2.
\end{equation}The total free energy cost on a surface of area $A$ is \begin{equation}
F = \frac{A}{a_{\mathrm{I}}^2}T \log 2
\end{equation}implying that the coefficient of surface tension $\gamma$ is \begin{equation}
\gamma = \frac{F}{A} \sim \frac{T\log 2}{a_{\mathrm{I}}^2} \sim 0.03 \; \frac{\mathrm{N}}{\mathrm{m}}.
\end{equation}Typical values for $\gamma$ at an interface between oil and water is $0.05$ N/m \cite{Phillips2012}, quite close to what we got with this very simple estimate.   This is the correct order of magnitude for \emph{many} liquid-air interfaces as well, suggesting that free energy penalties are responsible for surface tension.

There is a subtlety -- this derivation suggests that the surface tension would increase with temperature;  but the opposite is true \cite{Adam1941}.     A more complicated statistical derivation would correct for the oversight.

\subsection{Specific Heat}
The specific heat of gases and solids are very famous topics in statistical mechanics \cite{Reif1965}, sometimes even in introductory courses.   What is rather strange, however, is that such books do not point out that the specific heat of \emph{liquids} is also morally very similar, to within a factor of 2.  \cite{Bolmatov2012} has argued that quantitatively, the specific heat of liquids can be described by similar techniques.   Here, however, we will content ourselves with simply pointing out that the qualitative similarity of the specific heat between \emph{all} the classical phases of matter should not be a surprise.   This follows from the fact that \begin{equation}
c = \frac{1}{N} \frac{\mathrm{d}E}{\mathrm{d}T} \label{eqc}
\end{equation}where $N$ is the number of molecules.   We will be sloppy about what thermodynamic ensemble we are referring to (is $V$ or $P$ held constant?) as this is an O(1) effect.    If we work at fixed volume, then all exchanges of energy are due to heat transfer, and it is reasonable to expect heuristically that $\mathrm{d}E/\mathrm{d}T \sim \mathrm{d}(TS)/\mathrm{d}T$ is proportional to the entropy per molecule, $S/N \equiv \log \nu$.   Now, $\nu$ counts the effective number of degrees of freedom per particle at temperature $T$ -- we have argued, e.g., that $\nu\sim 6$ for water earlier.     It is more customary to talk about specific heat per mass (this is analogous to the fact that we want to scale out the number of atoms in each molecule).   In this case, we find, if $m$ is the mass of the molecule: \begin{equation}
c_m = \frac{c}{m} \sim \frac{\log \nu}{m}.
\end{equation}Heuristically, we expect (for very large, complex molecules such as polymers) $\nu \sim \exp[m/m_{\mathrm{H}}]$, as there is some flexibility in the orientation of each individual atom, so this ratio should qualitatively be constant among all materials.   This is indeed true to within an order of magnitude (with some exceptions like hydrogen gas) \cite{Atkins2009}:   the value turns out to be about 0.01--$1 / m_{\mathrm{H}}$ for almost all materials, including gases, at room temperature.   The fact that the mass of the molecule drops out of our estimate (and, for complex materials, out of the actual value of $c_m$) implies that $c_m$, as opposed to $c$, is the ``correct" unit for specific heat.

\section{A Non-Universal Result:  Evaporation Rates}
The calculation of evaporation rate \cite{Ma1985} is a rather famous problem, although as with many other calculations it is left in terms of many seemingly non-universal constants.    The typical way to understand the evaporation rate is to consider a liquid which is in chemical equilibrium with its vapor.   We can then determine the evaporation rate per unit area $\lambda$ by studying the rate at which vapor molecules enter the liquid.    This will be a simple example of a highly non-universal rate that we can derive;  in general, these are quite complicated.

In an introductory class, this can be done by dimensional analysis.   If the number density of molecules is of the vapor is $n$, the temperature of the gas is $T$, and the mass of the molecules is $m$, one finds that the rate $\mu$ at which vapor molecules enter the liquid is\begin{equation}
\mu \sim n\sqrt{\frac{T}{m}} 
\end{equation}Alternatively, we may want to use the ideal gas law $P=nT$ and express things in terms of a pressure at the surface of the liquid.   We will see from the liquid perspective that this is more natural.    

Now, we want to consider the evaporation rate $\lambda$ at which molecules leave the liquid.   Since the two systems are in chemical equilibrium, we have that \begin{equation}
\frac{\lambda}{\mu} = \mathrm{e}^{-E_{\mathrm{I}}/T}.
\end{equation}Now, the evaporation rate is usually expressed in terms of a saturation pressure \cite{Ma1985}, which must be determined in terms of microscopic, non-thermal properties of the liquid.   We have seen the natural scale of pressures in a liquid is $P_{\mathrm{sat}} \sim 1/\kappa$, as in Eq. (\ref{kappa}).   For water, one finds $P_{\mathrm{sat}}\sim 10^{10}$ Pa,\footnote{To obtain this result from empirical data, take the $T\rightarrow \infty$ limit of the Antoine equation for water (empirical form of Clausius-Clapeyron equation) \cite{Haynes2013}.} which is not too hard to imagine, as water is a fairly strongly interacting liquid.    We conclude that the evaporation rate should be \begin{equation}
\lambda \sim \frac{E_{\mathrm{I}}}{a^3_{\mathrm{I}}\sqrt{mT}} \mathrm{e}^{-E_{\mathrm{I}}/T}.
\end{equation}

This equation is the only example of a rather non-universal result that we will derive.   Here, it is easy to understand why:   there is a factor of $\mathrm{e}^{-E_{\mathrm{I}}/T}$, which is sensitive to a ratio of two energy scales.    These thermal activation factors are responsible for the rich variety of reaction kinetics in chemistry and biology, which allow for dynamics separated across large time scales, despite the fact that all energies are of similar order of magnitude.

Are there any other (non-electrical, non-chemical kinetic) material properties which cannot be obtained by dimensional analysis?   The answer is yes:  a famous example of a quantity which varies enormously between liquids is \emph{viscosity} \cite{Viswanath2007}.   The reason is that viscosity depends on microscopic interaction time scales which turn out to depend in great detail on the particular liquid being studied:  a dramatic demonstration of this is in supercooled glass-forming liquids \cite{Ediger2000}.

\section{Conclusion}
In this paper, we have presented a variety of very elementary estimations for the material properties of liquids and solids.   Most, if not all of them, are suitable for typical freshman introductory physics courses.   As we have already mentioned,  these estimates are so simple because almost all materials are made up of similar atomic and molecular building blocks;  it then follows from dimensional analysis that we expect a huge variety of material properties to appear rather universal, up to O(1) constants (which often are not larger than 10).     While these factors play hugely important roles in engineering and practical applications, it is important to understand how the fundamental scales of material properties are set.     An interesting (and increasingly important) exception to the universal orders of magnitude are cellular solids \cite{Courtney2000} or foamy liquids \cite{Weaire2001}, often made up of a ``mixture" of a liquid or solid with a gas;  large separations of scales between the properties of the two phases may lead to exotic material properties.

   It is worth suggesting to students that they can make estimates of molecular mass, energy and length scales using physics that was understood over a century ago.   For example:  by measuring the specific heat of water, we can estimate the degrees of freedom per mass (and thus, the mass of the water molecule);  the mass density then gives us the microscopic length scale, and the speed of sound gives us the microscopic energy scale.\footnote{This ``historical estimate" requires an understanding of the conversion between temperature to energy, first experimentally determined by Planck.  Because molecular physics is described by quantum mechanics, one of the scales must be found by recourse to this (or a quantum mechanical) argument.}

If this material is taught in a course, there are a couple of extensions that students should  explore on problem sets.  Students may be asked to derive some of the estimates in this paper for themselves, and look up the empirical values to see how their estimates hold up.   Some other types of sample questions for students are provided in an appendix.

\section*{Acknowledgements}\addcontentsline{toc}{section}{Acknowledgements}
We would like to thank Bruno Eckhardt and Jacob Sanders for helpful comments.   A.L.  is supported by the Smith Family Graduate Science and Engineering Fellowship.   

\begin{appendix}
\titleformat{\section}
  {\gdef\sectionlabel{}
   \Large\bfseries\scshape}
  {\gdef\sectionlabel{\thesection. }}{0pt}
  {\begin{tikzpicture}[remember picture,overlay]
	\draw (-0.2, 0) node[right] {\textsf{Appendix \sectionlabel#1}};
	\draw[thick] (0, -0.4) -- (\textwidth, -0.4);
       \end{tikzpicture}
  }
\titlespacing*{\section}{0pt}{15pt}{20pt}

\section{Questions for Students}
Here are some sample homework questions that can be provided to students, based on this material:

 \begin{enumerate}[\color{\stylecolor}\textsf{\textbf{\arabic{enumi}.}}]
 \item One of the lightest liquids known to humans is liquid helium:  $\rho\approx 100\; \mathrm{kg/m}^3$.   (By comparison, at room temperature, hydrocarbons are some of the lightest liquids, at $\rho\approx 600\; \mathrm{kg/m}^3$.) Make a crude estimate of the least dense possible liquid, and compare your answer to this empirical value.   
 \item Graphene is an effectively two dimensional compound of great interest in modern condensed matter physics.   Estimate a fundamental limit on the mass per surface area of any two dimensional compound made up of well-defined atoms.    In reality, graphene is arranged in a honeycomb arrangement:  each atom is connected to three neighboring atoms, as shown in Figure \ref{fig2}.   Modeling each carbon atom as a circle of radius 0.06 nm, and using $12m_{\mathrm{H}}$ as the mass of a carbon atom, estimate the mass per surface area.  Take into account the appropriate geometric factors.  Compare to the empirical value:  $\sigma \approx 7\times 10^{-7} \; \mathrm{kg/m}^2$.\footnote{Note to teachers:  for an introductory course, asking a similar question about a three dimensional material with any non-simple cubic crystal structure is unnecessarily challenging mathematically, for no extra gain in physical insight.}
  \begin{figure}[h]
\centering
\includegraphics{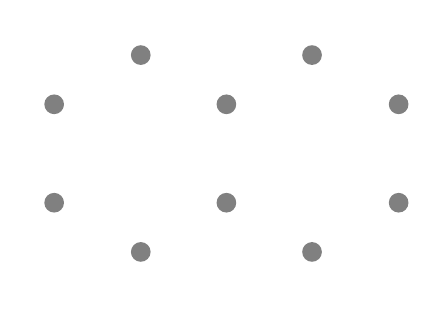}
\caption{The honeycomb lattice of graphene.}
\label{fig2}
\end{figure}
\item If the relevant length and energy scales are both intermolecular, estimate the Young's modulus of a solid.    Do you think there are type(s) of materials where these are the appropriate scales?   If so, look up the empirical values of $E$ for these materials, and see if your guess is correct.
\item A very interesting class of materials consists of \emph{cellular solids}:  for example, an array of tiny metal rods, with air filling the space between them.    Suppose a fraction $f$ of the material is made up of the metal rods.   What would the mass density of such a material be?   How would its Young's modulus compare to that of the pure metal?  Give your answers in terms of $f$, and any appropriate material properties. (These materials have very interesting nonlinear behavior due to the crumpling of the rods under a lot of stress.   The associated crumpling absorbs substantial energy -- thus, these materials are often used as protective packaging.)
 \item If electrons in atoms were replaced with muons,  the atomic energy scale $E_{\mathrm{H}}$ would be $10^{-16}$ J, the length scale would be $10^{-12}$ m, and the mass scale would be $10^{-27}$ kg.    Estimate the mass density, Young's modulus and speed of sound of a material made up muonic atoms, and compare to the values for materials made up of electronic atoms.
\item It is experimentally known that the melting temperature of a solid, $T_{\mathrm{m}}$, is related to its Young's modulus $E$ by some power:   $T_{\mathrm{m}} \sim E^\nu$.   What should the value of $\nu$ be?     Compare $T_{\mathrm{m}}$ with the energy scale of chemical bonds (you need to look up empirical data for some materials here) -- does $T_{\mathrm{m}}$ provide a good measure of the binding energy of a solid?   
\item If we model chemical bonds as harmonic oscillators (Hookean springs), what is the form of $V(u)$.    What would the value of $\alpha$ be?  Comment on the answer.  How would \emph{fluctuations} in the value of $u$ scale with $T$ and $E_{\mathrm{H}}$?  (It may be helpful to note that the typical deviation in $E_{\mathrm{bond}}$ will be of order $T$.   You do not need to do any integrals to solve this problem.) 
\item Can you provide a very simple argument why $\alpha \sim 1/T_{\mathrm{m}}$ (without referring to any molecular physics)?
\item Suppose the surface tension at the air-water interface was caused by intermolecular forces holding water together.    Estimate $\gamma$ in this case.   The experimental value of the surface tension is about $\gamma \approx 0.06$ N/m.    Is our hypothesis possible?   Compare to the hypothesis that free energy penalties are responsible for $\gamma$.
 \item Should the specific heat (measured in experimental units of specific heat per mass) of lithium metal be large or small compared to uranium metal?   Estimate the ratio $c_{m,\mathrm{Li}}/c_{m,\mathrm{U}}$.
\item A shear stress $\sigma_{xy}$ emerges when a viscous fluid is placed under a transverse velocity gradient.  In particular, the viscosity $\eta$ is defined by the relation $\sigma_{xy} = \eta \times \partial v_x/\partial y$.    What are the units of $\eta$?   Using the mass of a water molecule, the intermolecular energy, and the intermolecular spacing, estimate by dimensional analysis the viscosity of liquid water.   The empirical value is about $10^{-3}$ (in the appropriate SI units).   Compare your answer to this.   The viscosity of honey (made up of organic molecules) is about 1000 times larger than water.  Which of your estimates of molecular scales may be incorrect?  Viscosity is very hard to estimate theoretically -- can you think of a reason why?
 \end{enumerate}

\end{appendix}

\bibliographystyle{unsrt}
\addcontentsline{toc}{section}{References}
\bibliography{macrobib}

\end{document}